%% file: p670.tex
\documentclass[10pt,conference,letterpaper]{IEEEtran}
\usepackage{times,amsmath,epsfig}
\usepackage{subcaption}
\usepackage{xcolor}

\title{QR2: A Third-party Query Reranking Service Over Web Databases}

\author{%
	{
		Yeshwanth D. Gunasekaran{\small $~^{\dag}$}, 
		Abolfazl Asudeh{\small $~^{\dag\dag}$},
		Sona Hasani{\small $~^{\dag}$},
		Nan Zhang{\small $~^{\ddag}$},
		Ali Jaoua{\small $~^{\dag\ddag}$},
		Gautam Das{\small $~^{\dag}$}
	}%
	\vspace{1.6mm}\\
	\fontsize{10}{10}\selectfont\itshape
	$~^{\dag}$University of Texas at Arlington, $~^{\dag\dag}$University of Michigan, $~^{\ddag}$Pennsylvania State University, $~^{\dag\ddag}$Qatar University\\
	\fontsize{9}{9}\selectfont\ttfamily\upshape
	
}

\begin{document}
\maketitle
\begin{abstract} 
The ranked retrieval model has rapidly become the de-facto way for search query processing in web databases.  Despite the extensive efforts on designing better ranking mechanisms, in practice, many such databases fail to address the diverse and sometimes contradicting preferences of users.
In this paper, we present QR2, a third-party service that uses nothing but the public search interface of a web database and enables the on-the-fly processing of queries with any user-specified ranking functions, no matter if the ranking function is supported by the database or not.
\end{abstract}
\input{intro}

\input{QR2system}
\input{demo}

\vspace{-2mm}
\input{summary}

\vspace{-3mm}
\section{Acknowledgement}
This contribution was made possible by NPRP grant No. 07-794-1-145 from the Qatar National Research Fund (a member of Qatar Foundation). Any findings, conclusions, or recommendations expressed in this material are those of the authors and do not necessarily reflect the views of the sponsors listed above.

\bibliographystyle{IEEEtran}
\bibliography{ref}
\end{document}

%% file: intro.tex
\section{Introduction}\label{sec:intro}

The ranked retrieval model~\cite{PREFER, chomicki} has rapidly become the de-facto way for query processing in web databases.
Instead of returning all of the search query matches, the ranked retrieval model orders the matching tuples according to an often proprietary ranking function, and returns the top-$k$~\cite{fagin2003}.
This model is a natural fit for the web databases, as the short attention span of web users demands the most desirable tuples to be returned first and, in addition, achieving a short response time requires to limit the length of returned results to a small value.
Nevertheless, this puts more responsibilities on the web database designer, as the ranking mechanism must properly capture the need of database users. In practice, web users often have diverse and sometimes contradicting preferences on numerous factors while many web databases may not design the most effective ranking functions to reflect them.
Examples of such designs range from rental and real estate websites such as Zillow\footnote{\footnotesize{www.zillow.com}} to shopping websites such as Blue Nile\footnote{\scriptsize{www.bluenile.com}}, the largest diamonds online retailer, for ranking examples of price per square feet for Zillow and the aggregation of depth and table percent for Blue Nile. As a result, there is often a significant gap, in terms of both design and diversity, between the ranking function(s) supported by the web database and the true preferences of the database users.

While ranked retrieval model~\cite{PREFER, chomicki}, top-$k$ query processing~\cite{fagin2003, das2006views}, and building indices for them~\cite{chang2000onion, regretratio}
on one side, and hidden web databases~\cite{madhavan2008google, sheng2012optimal, lu2015hidden} on the other side
are well studied in the literature, 
research on ranked retrieval model on web databases is limited to the two recent papers~\cite{skylinediscovery, queryreranking}.
Similar to~\cite{queryreranking},~\cite{mobiface} also proposes a third-party service on web databases, however unlike~\cite{queryreranking}, it focuses on faceted search for mobile applications.
In this paper, we demonstrate QR2, a working solution based on \cite{queryreranking}, a third-party service that uses nothing but the public search interface of a web database to enable the on-the-fly processing of queries with any user-specified ranking functions.
Proposed algorithms in \cite{queryreranking} enable the ``Get-Next'' primitive that provide an incremental reranking of the results.

\noindent\textbf{Problem definition}: Consider a web database $D$ with a top-$k$ interface and an arbitrary, unknown, system ranking function. Given a user query $q$, a user-specified monotonic ranking function $f$, and the top-$h$ ($h \geq 0$ can be greater than, equal to, or smaller than $k$) tuples satisfying $q$ according to $f$, discover the No.~$(h+1)$ tuple for $q$ while minimizing the number of queries issued to the database $D$.

\noindent\textbf{System Overview}: 
Proposed solutions address the problem in two settings: (i) 1D, where the user-desired ranking function is on a single attribute, and (ii) MD, the ranking of two or more attributes.
In this paper, we consider a linear combination of the attribute values of each tuple as the score of it. The tuples are ranked based on this score.
The algorithms are based on the principle of covering a region of interest specified by the so-called rank-contour of the best-known solution.
Generally speaking, for each setting the following algorithms are proposed\footnote{\footnotesize For MD, {\em MD-TA}, an implementation of TA~\cite{fagin2003} using 1D-RERANK, is also proposed.} (please refer to~\cite{queryreranking} for more details):
\begin{itemize}[
    \setlength{\IEEElabelindent}{\dimexpr-\labelwidth-\labelsep}
    \setlength{\itemindent}{\dimexpr\labelwidth+\labelsep}
    \setlength{\listparindent}{\parindent}
  ]
\item {\em  (1D/MD)-BASELINE}:
In high-level, 1D and MD baseline algorithms start by the broad queries that cover the search space and, using the best-known tuple as the upper-bound, keep narrowing the search space until the top tuple is found.
\item {\em  (1D/MD)-BINARY}: 
In certain cases, especially when the user-specified ranking function is anti-correlated with the ranking function of the web database, baseline algorithms have a poor performance. Hence, rather than issuing the query on the whole search space, (1D/MD)-BINARY applies a binary search on the space to reduce its size by half at every iteration.
The binary algorithms, however, perform badly in dense regions, i.e., when a large number of tuples are clustered together within a small interval.
\item {\em  (1D/MD)-RERANK}:
In order to resolve the problem of the dense regions in (1D/MD)-BINARY algorithms, (1D/MD)-RERANK applies the idea of on-the-fly indexing of these regions. i.e., if the density of the region of interest is more than a threshold, it directly gets the top tuple from an oracle that indexes these regions as needed.
\end{itemize}

\noindent\textbf{Web Databases}:
In this demonstration, we provide the reranking service on two major web databases from two different areas, namely Blue Nile and Zillow. 
As a well-known real-estate web database with millions of entities, Zillow allows us to showcase the performance of the system on a large database.
Blue Nile, on the other hand, is the biggest online diamond retailer. What makes Blue Nile interesting in this demonstration is that diamonds have more ranking features such as \texttt{\small Carat}, \texttt{\small Depth}, and \texttt{\small Table} that can show the performance of the system in high dimensions.

%% file: QR2system.tex
\section{QR2 system}\label{sec:QR2system}
In this section we describe QR2's system architecture, its technical challenges, and the user interface.
\subsection{Architecture}\label{subsec:arch}

\begin{figure}[t]
    \centering
    \includegraphics[width=0.49\textwidth]{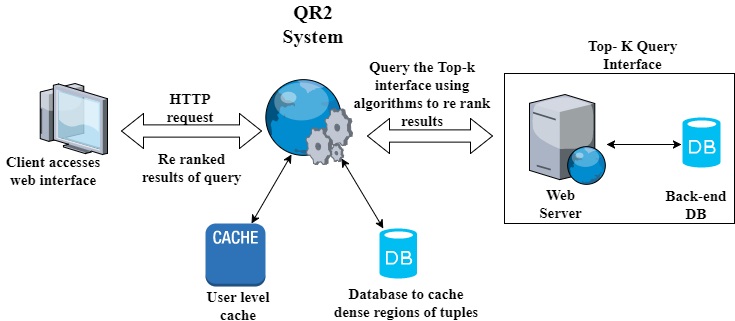}
    \vspace{-2mm}
    \caption{\small{Architecture of QR2.}}
    \vspace{-6mm}
    \label{fig:architecture}
\end{figure}
Fig.~\ref{fig:architecture} shows the  architecture of QR2.
Web service is the central component of the architecture, where the users connect via Internet and select the data source. 
Once a user submits a query along with a ranking preference, the server creates a new session and processes the user’s request. The session variable (user level cache) is used to store the tuples that are already "seen" while discovering the top-$h$ of the given query, in order to accelerate the query processing and subsequent get-next operations.
In addition to the query history, retained in the session variable, 1D-RERANK and MD-RERANK apply an on-the-fly indexing that detects the dense regions and pro-actively crawls top ranked tuples to save on processing future user queries. We use a MYSQL database to store the tuples in the dense region.

\input{tech_challenges}

\begin{figure*}[!ht]
    \centering
    \includegraphics[width=1\textwidth]{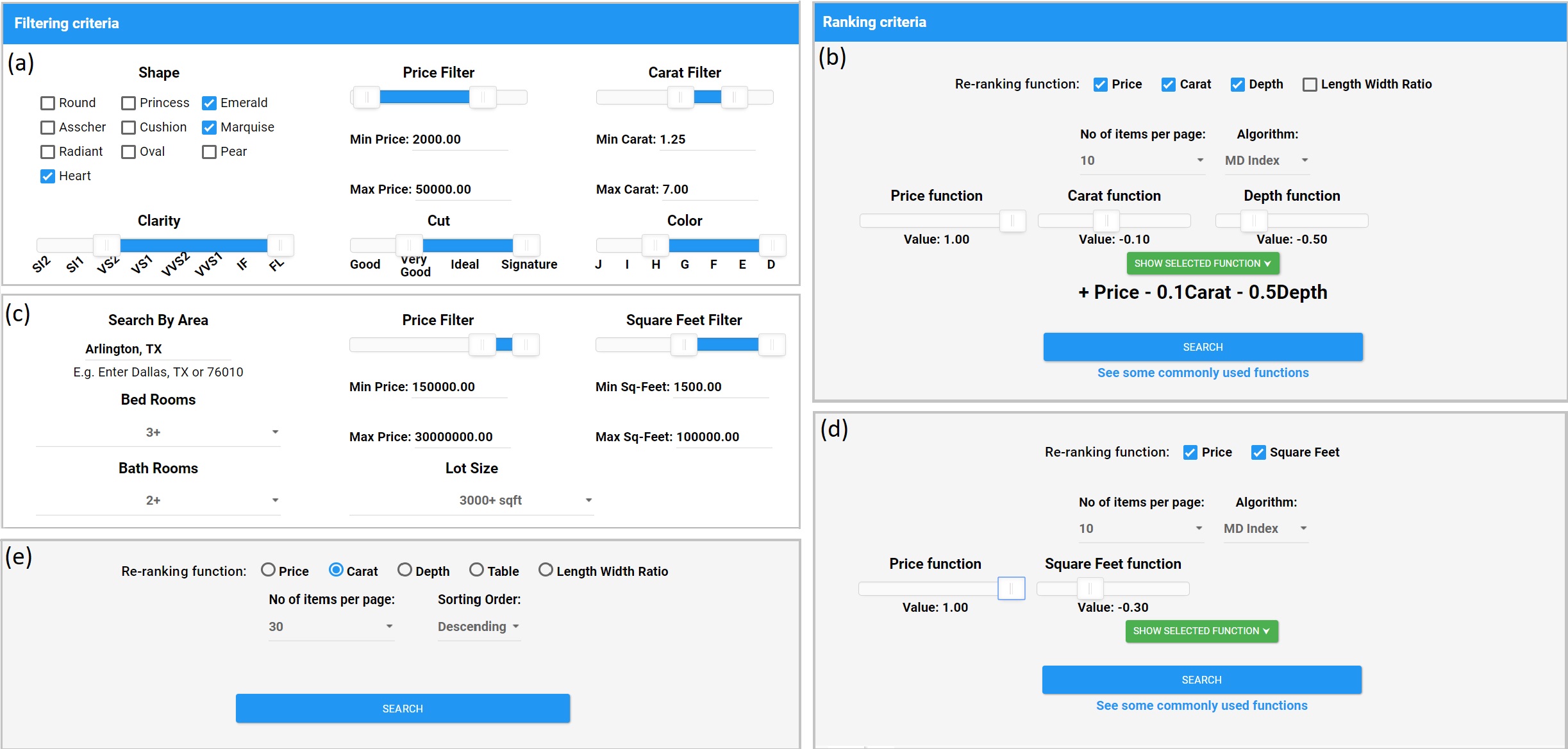}
    \vspace{-2mm}
    \caption{\small User interface of QR2}
    \vspace{-6mm}
    \label{fig:fil_and_rank}
\end{figure*}

\subsection{User Interface}\label{subsec:userInterface}
QR2 has three main sections in its user interface, (i) Filtering section, (ii) Ranking section, and (iii) Search results. 
Below is the description of these sections:

\noindent
\textbf{Filtering section:} The filtering section is used for specifying filtering predicates using a user-friendly web interface. This interface is common in many web databases, especially in Blue Nile and Zillow. For Blue Nile, the user can adjust the \texttt{\small price}, \texttt{\small carat}, \texttt{\small cut}, \texttt{\small color}, and \texttt{\small  clarity} sliders to search between a particular area and also select nominal attributes like \texttt{\small shape} of diamond.
For Zillow, in addition to the location (e.g. the city and zip code), as shown in Fig.~\ref{fig:fil_and_rank}(c) and (d), we include all the filtering conditions, such as \texttt{\small number of bedrooms} and \texttt{\small price}, in the filtering section.

\noindent\textbf{Ranking section:}
For QR2, as a reranking service, the ranking section is special.
The ranking section should provide a user-friendly way of identifying the user preference, even for the users that do not have an understanding of the ranking function notion. Obviously, expecting the user to compose a function for the query is not realistic.
Hence, one of the challenges of this project was designing this section in a way that is convenient for the ordinary users.
After investigating different alternatives, we designed the ranking section as follows:
\begin{itemize}[
    \setlength{\IEEElabelindent}{\dimexpr-\labelwidth-\labelsep}
    \setlength{\itemindent}{\dimexpr\labelwidth+\labelsep}
    \setlength{\listparindent}{\parindent}
  ]
\item {\em 1D}: Similar to the {\em order by} clause in a SQL query, for one-dimensional reranking, the user needs to simply specify the ranking attribute, as well the ordering direction, i.e., ascending or descending. As specified in Fig.~\ref{fig:fil_and_rank}(e), in addition to the ranking attribute and the sorting direction, the user can specify the number of results per page.

\item 
{\em MD}: This component aims to provide a convenient way for the ordinary users to specify their preference.
To do so, after normalizing the attribute domains, it uses a slider for each of attributes chosen for ranking. For each attribute $A_i$, the preference co-efficient, $w_i$ is specified by a slider value in the range of [-1,1].
Based on the slider values, the user-specified ranking function is $\sum\limits_{i=1}^m w_i.A_i$.
Fig.~\ref{fig:fil_and_rank}(c) and Fig.~\ref{fig:fil_and_rank}(d) show two example instances of the MD ranking section for Blue Nile and Zillow, for the ranking functions \texttt{\small price - 0.1 carat - 0.5 depth} and \texttt{\small price- 0.3 square feet}, respectively.
In addition to the slider, we also suggest a list of popular functions for the user to choose from.
\end{itemize}

\begin{figure}[!ht]
	\centering
	\includegraphics[width=0.48\textwidth]{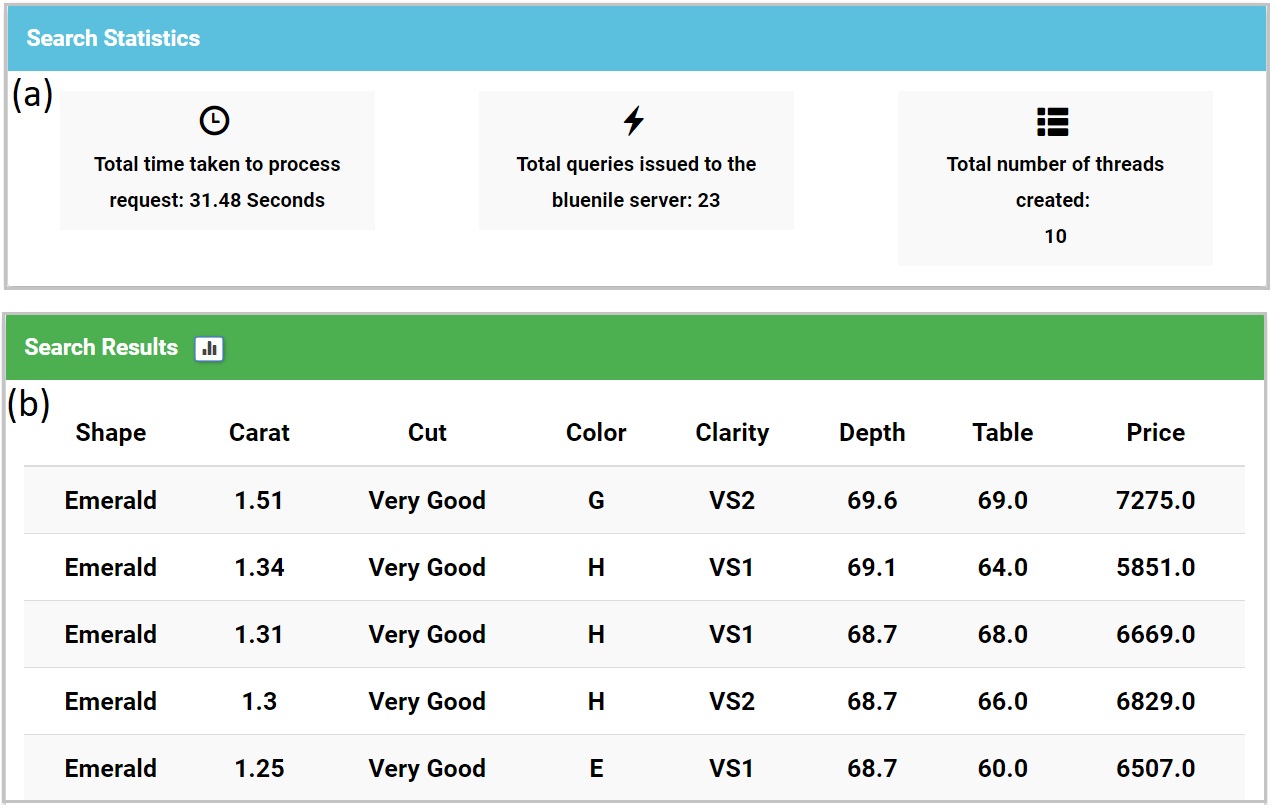}
	\vspace{-2mm}
	\caption{\small Search results and statistics of QR2}
	\vspace{-2mm}
	\label{fig:md_bluenile_results}
\end{figure}

\noindent\textbf{Search results and statistics:}
Once the query is issued, the system processes and returns the table of top-$k$ results ($k$ is specified by the user). 
Each row of the table shows the details of a tuple while clicking on it opens the web database page of the tuple.
The get-next button allows the user to get the next page of results. 
Along with the results, the user is also provided with a small panel providing statistics such as
query cost in terms of the number of queries issued to the web database and processing time. Fig.~\ref{fig:md_bluenile_results} shows a screen shot of results and the statistics for a reranking query on Blue Nile. Similarly, for Zillow and the ranking function Price - 0.3*Carat, the system
issued 27 queries to the Zillow server, which took 33 seconds.

%% file: tech_challenges.tex
\subsection{Technical Challenges}\label{subsec:technicalChallenges}
\begin{figure}[t]
	\centering
	\includegraphics[width=0.45\textwidth]{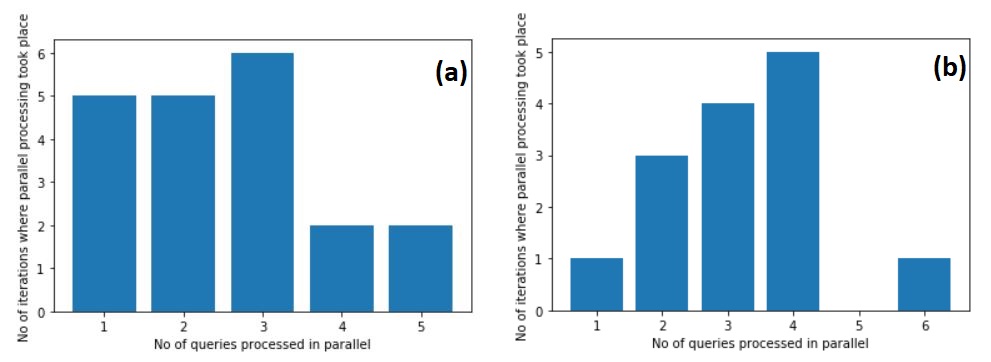}
	\vspace{-2mm}
	\caption{\small{Parallel processed queries per iteration for (a) three and (b) two dimensions (BlueNile)}}
	\vspace{-6mm}
	\label{fig:limit_parallel_process}
\end{figure}

In addition to the technical challenges addressed in~\cite{queryreranking}, the following are the main challenges we faced in practice, together with their resolutions:

\begin{itemize}[
    \setlength{\IEEElabelindent}{\dimexpr-\labelwidth-\labelsep}
    \setlength{\itemindent}{\dimexpr\labelwidth+\labelsep}
    \setlength{\listparindent}{\parindent}
  ]    
\item {\em General positioning assumption}: a general assumption in~\cite{queryreranking} is that
no two tuples have the same values on a given attribute.
This assumption, however, may not hold in practice. Especially when the number of tuples matching the predicate $t[A_i] = V_c$ is greater than system-$k$, the issued query to the web database never underflows.
To solve this, we implemented the crawling algorithm proposed in~\cite{sheng2012optimal}. QR2 calls this function when 
the number of tuples matching a value $V_c$ is greater than system-$k$.

\item {\em Attributes with different cardinalities}: 
Handling the attributes with different domains is left as a part of ranking function design in~\cite{queryreranking}.
However, in practice, it does not seem realistic to expect the users to take the burden. 
Thus, we apply the min-max normalization of attributes values to resolve this issue. Please note that obtaining the min and max values on each attribute is simply doable using the 1D-RERANK algorithm.

\item {\em Parallel processing}:
QR2, as a third party service, may have different users issuing different queries at the same time; therefore, the sequential processing of queries may significantly reduce the system performance. In addition to a non-sequential processing of different queries, we apply parallel processing while performing each query, in order to reduce the query processing time. We note that this may, sometimes, increase the number of queries issued to the web database. Specially, 
the following parallel processings help reducing the effect of the web database delay: 
\begin{itemize}
\item In order to verify that the top discovered tuple is indeed the true top one, we issue several queries, in parallel, that cover the areas in which a tuple may dominate the discovered tuple. 
\item In MD, after the initial get-next, in order to discover subsequent tuple(s), the algorithm partitions the search space on an attribute $A_i$ and searches the two subspaces independently. The subsequent tuple is the top tuple from these two regions with the best score. Since the search in subspaces is done independently, it is easily parallelable.
\end{itemize}
\noindent

Fig.~\ref{fig:limit_parallel_process} shows the number of iterations where parallel processing took place in an experiment on Blue Nile for both two and three dimensional searches.
One can see that in the 3D experiment (Fig.~\ref{fig:limit_parallel_process}-a) more than 90\% of queries were submitted in parallel. Similarly, for the 2D search (Fig.~\ref{fig:limit_parallel_process}-b), only one out of 45 queries issued sequentially -- i.e., more than 97\% percent of queries issued in parallel.

\item {\em Managing the dense region cache}:
As stated earlier, 1D-RERANK and MD-RERANK apply on-the-fly indexing of the dense regions to speed up the future processing.
Being shared between all the users, the index may become relatively large, not to fit in the main memory. Thus, we use 
MYSQL to store the dense regions. Additionally, before the system boots up we verify the cache and update the changes from the web database.
\end{itemize}

%% file: demo.tex
\section{Demonstration Plan}\label{sec:demo}
In this section, we describe different scenarios that the user can use to interact with QR2 and provide a few more specific case studies for Blue Nile and Zillow.

\subsection{System Implementation:} The QR2's back-end is implemented using Python 3.5. To build the web service and the web interface we used Flask framework due to its minimal footprint, session management, and database connection tools. We use Requests library for query processing and Pandas library to save the query results in data frames. Pandasql, a library which enables SQL queries over pandas data frames, is used to facilitate the query processing phase. On the client side, Javascript, HTML, and CSS are used to parse the results and show them to the user.



\subsection{Demonstration Scenarios:}
In order to study the performance of different algorithms, we consider different combinations of (i) both web databases, (ii) 1D and MD algorithms, (iii)  filtering conditions, and more importantly (iv) ranking functions that are independent, positively correlated, and negatively correlated with the web database's system ranking function:

\noindent{\bf 1D}: The reranking, in this case, is on a single attribute. For both Zillow and Blue Nile and for queries with different filtering predicates, we will choose different attributes for ranking. Also, to construct the rankings with different correlations with the system ranking function, we will test the performance of algorithms in both ascending and descending orders.

\noindent\textbf{MD:}
The MD reranking is on more than one attribute, where 
the user-specified ranking function is the dot product of the slider values with the ranking attributes. In order to construct queries with different correlations with the system ranking function, we test different combinations of positive and negative slider values on different numbers of attributes. Especially, we choose Blue Nile for constructing ranking functions with more than two ranking attributes. Fig.~\ref{fig:fil_and_rank}(b) shows an example of such ranking functions (\texttt{\small price - 0.1 carat - 0.5 depth}).

    
\noindent\textbf{On-the-fly indexing:} 
Indexing the dense regions for future is the main technique used in 1D-RERANK and MD-RERANK to resolve the performance issues of both (1D/MD)-BASELINE and (1D/MD)-BINARY. Showing the effectiveness of this technique is part of the demonstration plan. To do so, after issuing multiple queries, we will track the performance of (1D/MD)-RERANK in terms of both processing time and the number of submitted queries to the web database.
       
\noindent\textbf{Best v.s. worse cases:}
Finally, we will demonstrate some of the best and worst case scenarios to show efficiency and limitations of the system.
For example, we will show that when a large number of tuples have the same value $V$ on an attribute $A_i$, 
the performance of the system may drop significantly.
That is because,
in order to identify the next top tuple, the system may first need to crawl all tuples where $t[A_i] = V$.
On the other hand, 
when the attribute values follow a uniform distribution on the domain space, even the binary search strategy performs well.
Here are two of such functions:
\begin{itemize}[
    \setlength{\IEEElabelindent}{\dimexpr-\labelwidth-\labelsep}
    \setlength{\itemindent}{\dimexpr\labelwidth+\labelsep}
    \setlength{\listparindent}{\parindent}
  ] 
    \item The function \texttt{\small price + LengthWidthRatio} is inefficient to run on Blue Nile.
    While processing this query, QR2 needs to crawl all the tuples with \texttt{\small t[LengthWidthRatio] = 1}. In Blue Nile, when writing this paper, around 20\% of the tuples satisfy this predicate. The system, therefore, needs to crawl all these tuples before returning the results.
    Note that thanks to the on-the-flying indexing, (1D/MD)-RERANK will still have a low amortized cost in these cases.
    \item The function \texttt{\small price + squarefeet} runs fast on Zillow.
     The goal of this function is to find the houses with low price and small square feet.
     The positive correlation between attributes \texttt{\small price} and \texttt{\small squarefeet}, as well as the positive correlation of this query with Zillow's system ranking function, makes the algorithms to finish quickly.     
\end{itemize}

%% file: summary.tex
\section{Summary}\label{sec:summary}
We proposed to demonstrate QR2, a third party service that enables the on-the-fly processing of queries with any ranking function defined by the user to a web database. Our system uses nothing but the public search interface of the web database and addresses a wide range of users preferences in ranking the results, even if not supported by the database.

%% file: p670.bbl
\begin{thebibliography}{10}
\providecommand{\url}[1]{#1}
\csname url@samestyle\endcsname
\providecommand{\newblock}{\relax}
\providecommand{\bibinfo}[2]{#2}
\providecommand{\BIBentrySTDinterwordspacing}{\spaceskip=0pt\relax}
\providecommand{\BIBentryALTinterwordstretchfactor}{4}
\providecommand{\BIBentryALTinterwordspacing}{\spaceskip=\fontdimen2\font plus
\BIBentryALTinterwordstretchfactor\fontdimen3\font minus
  \fontdimen4\font\relax}
\providecommand{\BIBforeignlanguage}[2]{{%
\expandafter\ifx\csname l@#1\endcsname\relax
\typeout{** WARNING: IEEEtran.bst: No hyphenation pattern has been}%
\typeout{** loaded for the language `#1'. Using the pattern for}%
\typeout{** the default language instead.}%
\else
\language=\csname l@#1\endcsname
\fi
#2}}
\providecommand{\BIBdecl}{\relax}
\BIBdecl

\bibitem{PREFER}
V.~Hristidis and Y.~Papakonstantinou, ``Algorithms and applications for
  answering ranked queries using ranked views,'' \emph{VLDB Journal}, 2004.

\bibitem{chomicki}
J.~Chomicki, ``Preference formulas in relational queries,'' \emph{TODS}, 2003.

\bibitem{fagin2003}
R.~Fagin, A.~Lotem, and M.~Naor, ``Optimal aggregation algorithms for
  middleware,'' \emph{Journal of Computer and System Sciences}.

\bibitem{das2006views}
G.~Das, D.~Gunopulos, N.~Koudas, and D.~Tsirogiannis, ``Answering top-k queries
  using views,'' in \emph{VLDB}, 2006.

\bibitem{chang2000onion}
Y.-C. Chang, L.~Bergman, V.~Castelli, C.-S. Li, M.-L. Lo, and J.~R. Smith,
  ``The onion technique: indexing for linear optimization queries,'' in
  \emph{SIGMOD}, 2000.

\bibitem{regretratio}
A.~Asudeh, A.~Nazi, N.~Zhang, and G.~Das, ``Efficient computation of
  regret-ratio minimizing set: A compact maxima representative,'' in
  \emph{SIGMOD}, 2017.

\bibitem{madhavan2008google}
J.~Madhavan, D.~Ko, {\L}.~Kot, V.~Ganapathy, A.~Rasmussen, and A.~Halevy,
  ``Google's deep web crawl,'' \emph{VLDB}, 2008.

\bibitem{sheng2012optimal}
C.~Sheng, N.~Zhang, Y.~Tao, and X.~Jin, ``Optimal algorithms for crawling a
  hidden database in the web,'' \emph{VLDB}, 2012.

\bibitem{lu2015hidden}
Y.~Lu, S.~Thirumuruganathan, N.~Zhang, and G.~Das, ``Hidden database research
  and analytics (hydra) system.'' \emph{IEEE Data Eng. Bull.}

\bibitem{skylinediscovery}
A.~Asudeh, S.~Thirumuruganathan, N.~Zhang, and G.~Das, ``Discovering the
  skyline of web databases,'' \emph{VLDB}, vol.~9, no.~7, pp. 600--611, 2016.

\bibitem{queryreranking}
A.~Asudeh, N.~Zhang, and G.~Das, ``Query reranking as a service,'' \emph{VLDB},
  vol.~9, no.~11, pp. 888--899, 2016.

\bibitem{mobiface}
A.~Nazi, A.~Asudeh, N.~Zhang, A.~Jaoua, and G.~Das, ``Mobiface: A mobile
  application for faceted search over hidden web databases,'' \emph{ICCA},
  2017.

\end{thebibliography}
